\documentclass{article}

\PassOptionsToPackage{numbers,compress}{natbib}

\usepackage[final]{tackling_climate_workshop_style}

\usepackage[utf8]{inputenc} 
\usepackage[T1]{fontenc}    
\usepackage{hyperref}       
\usepackage{url}            
\usepackage{booktabs}       
\usepackage{graphicx}       
\usepackage{amsmath}        
\usepackage{amsfonts}       
\usepackage{nicefrac}       
\usepackage{microtype}      

\title{Run it Back: Autoregression vs. Video Diffusion for Final-Conditioned Weather Trajectories}

\author{Jerry Lin \\
  Department of Computing and Data Sciences \\
  Boston University \\
  Boston, MA 02215 \\
  \texttt{jlin404@bu.edu}
  \And
  Elizabeth A. Barnes \\
  Department of Computing and Data Sciences \\
  Boston University \\
  Boston, MA 02215 \\
  \texttt{eabarnes@bu.edu}
}

\begin{document}

\maketitle

\begin{abstract}

Low-likelihood, high-impact extremes are projected to increase in severity and frequency in a warming climate. Sampling the precursors that lead to these events traditionally requires massive ensembles that scale poorly with lead time and rarity, and typical rare-event sampling and ensemble-boosting methods do not show how a single extreme might arise from diverse synoptic precursors. Recently, Lin et al.\ (2026) showed this is possible with final-conditioned 66-hour trajectories (trajectories constrained to end at an observed state) from cBottle-video, a pre-trained, off-the-shelf video diffusion climate emulator not explicitly fine-tuned for this task (Brenowitz et al., 2025). However, its ensembles were underdispersed and geostrophically inconsistent, and it reproduces conditioned frames only approximately, hindering autoregressive rollout. Here, we train Rewind-AR, a backward-in-time probabilistic autoregressive model with a FourCastNet~3 backbone (Bonev et al., 2025), and generate final-conditioned ensembles for the same three cases: the 2021 Pacific Northwest heatwave, Hurricane Ian, and Superstorm Sandy (in the Supplementary Information). The leading modes of Rewind-AR's antecedent 500 hPa geopotential height explain far more of the across-member variance than cBottle-video's for both hurricane position and heatwave surface temperature, and Rewind-AR better reproduces the day-over-day heatwave intensification observed in ERA5. Finally, geostrophic inconsistencies close to the conditioning frame seen in cBottle-video are not reproduced by Rewind-AR. Nevertheless, up to 22\% of Rewind-AR members become numerically unstable within 66 hours, so improving rollout physical consistency and stability over long lead times is a natural next step.

\end{abstract}

\section{Introduction}

With a warming climate, humanity is likely to experience more severe heatwaves, droughts, hurricanes, and floods \citep{Domeisen2022-gj,Hirabayashi2013-ry, Knutson2020-tj, Fischer2025-na, IPCC2023Sections, Cook2020-xw}. Understanding how such events can unfold is critical for risk assessment and scenario planning, but traditionally requires generating large physics-based ensembles at high computational cost. Because large ensembles scale poorly with lead time and event rarity, emerging techniques such as ensemble boosting, initial-condition optimization, and rare-event sampling directly target stronger extremes without exhaustive search \citep{Ragone2018-rj, Gessner2021-qy, Finkel2024-ln, Finkel2026-uu, Whittaker2026-fe, Hakim2026-oy, Lancelin2026-om}. While useful, none of these techniques attempt to sample the full distribution of trajectories that lead to the \textit{same} event. For local stakeholders, however, identifying stronger extremes in isolation is insufficient; contingency planning requires characterizing a range of plausible antecedent pathways that may culminate in a specific regional disaster.

Recently, \citet{Lin2026-sz} used Climate in a Bottle-Video (cBottle-video), a video diffusion climate emulator, to generate 1,000-member ensembles that converge on a single extreme event across three case studies---the 2021 Pacific Northwest heatwave, Superstorm Sandy, and Hurricane Ian---with antecedent 500 hPa geopotential height spread 84--89\% as large as that of start-conditioned ensembles. Because cBottle-video is trained using general-purpose masked conditioning over fixed 66-hour windows, this naturally raises the question: can a dedicated autoregressive architecture trained explicitly on reverse-time state transitions generate more physically credible backward trajectories? Here, we address this question by developing Rewind-AR, a backward-in-time probabilistic autoregressive model based on a FourCastNet~3 backbone \citep{Bonev2025-dw} and trained directly for reverse-time rollout. Although up to 22\% of Rewind-AR members become numerically unstable or unphysical within 66 hours, the leading modes of the surviving members' antecedent 500 hPa geopotential height ($z_{500}$) explain far more of the across-member variance than cBottle-video's over the same 66-hour window, both in the heatwave's surface temperature and in the hurricanes' positions. Nevertheless, for Superstorm Sandy (shown in the Supplementary Information), some of the trajectories imply implausible starting points, so more work remains to ensure that generated backward trajectories are actually representative of what could happen in reality.

\section{Methods}

\subsection{Rewind-AR}

Rewind-AR is a FourCastNet~3 backbone (273.5M parameters, counting each complex weight as two real parameters) configured as a probabilistic autoregressive model of the reverse-time conditional distribution $p(\mathbf{x}_{t-6\mathrm{h}}\mid\mathbf{x}_t)$ \citep{Bonev2025-dw}. The model is trained on 10 years of ERA5 reanalysis data (2009--2018, inclusive, with 2019 used for validation) on a regular $0.25^\circ$ latitude--longitude grid ($\sim$25\,km resolution) and generates multistep trajectories through autoregressive six-hour rollouts \citep{Hersbach2020-oh}. Full architecture, state representation, normalization, objective, and training details are provided in Supplementary Section~\ref{sec:methods_fcn3_backward}.

\subsection{cBottle-video}
\label{sec:methods_cbottle_main}

cBottle-video is a score-based video diffusion model trained on 38 years of ERA5 reanalysis data (1980--2017, validated on 2018) and four years of ICON data (2020--2024, validated on 2024--2025) regridded to HPX64, an equal-area HEALPix grid with 49,152 pixels across the globe ($\sim$100\,km resolution). It generates 66-hour (12-frame) trajectories simultaneously with the ability to arbitrarily condition on any subset of frames \citep{Brenowitz2025-mz, Lin2026-sz}. Case study results for this model are conditioned on the ERA5 dataset label and taken directly from \citet{Lin2026-sz}. Because cBottle-video regenerates its conditioning frames rather than copying them, it reproduces them only approximately \citep{Lin2026-sz}, whereas Rewind-AR starts from the conditioning state exactly. Full architecture, masked-conditioning formulation, and ODE sampling details are described in \citet{Brenowitz2025-mz} and in Supplementary Section~\ref{sec:methods_cbottle}.

\subsection{Case-study ensembles}

Each model generates 1,000 members per case. We call the conditioned final state the pinned frame and the state 66 hours prior the free end. Rewind-AR is initialized from the ERA5 state at the pinned frame and rolled backward in six-hour steps, and we analyze the first 66 hours (11 steps). Within an ensemble, all members share the same conditioning and differ only in their random draws. For cBottle-video, these are the sampler's initial latent noise. The model is inferenced with a batch size of one, and member $k$ ($k = 0, \dots, 999$) is seeded with $k$ \citep{Lin2026-sz}. For Rewind-AR, stochastic noise fields are supplied as extra input channels alongside the ERA5 state. The model is inferenced with a batch size of four, and 250 seeds are used to generate each 1,000-member ensemble.

Numerical stability screening (Rewind-AR only) excludes rollouts whose global $t_{2\text{m}}$ field becomes nonfinite or leaves $[150,350]$\,K, whereas track quality control (for Hurricane Ian and Superstorm Sandy) applies to both Rewind-AR and cBottle-video and requires an identifiable storm with a detected position at the free-end frame (Supplementary Section~\ref{sec:si-screening}).

\section{Results}

\subsection{Hurricane Ian}

\begin{figure}[t]
\centering
\includegraphics[width=\linewidth]{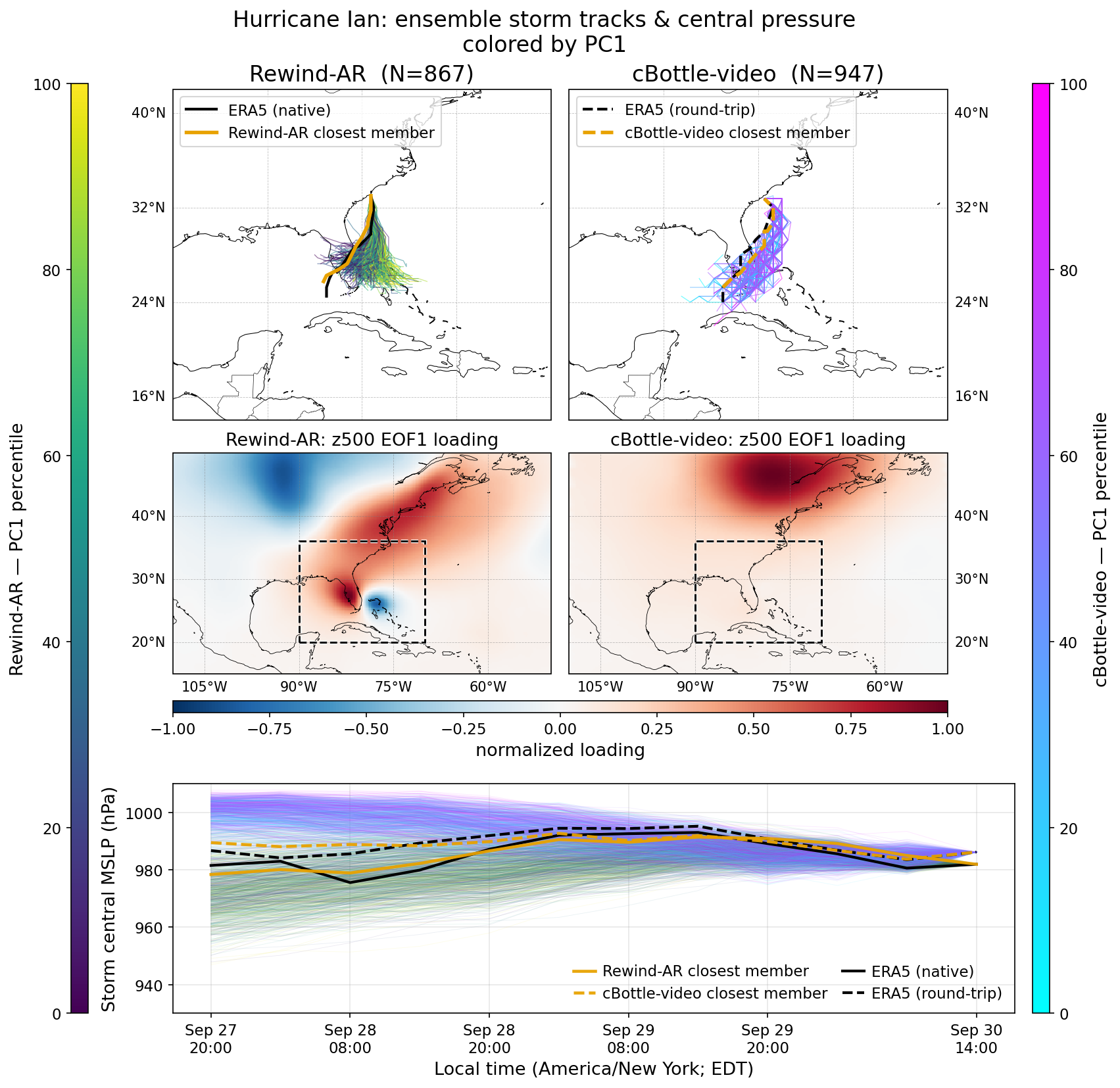}
\caption{Hurricane Ian 66-hour final-conditioned ensembles from Rewind-AR (left) and cBottle-video (right). (Top) Storm tracks identified using TempestExtremes and colored by free-end $z_{500}$ PC1 percentile \citep{Ullrich2021-nm, Lin2026-sz}. (Middle) Leading $z_{500}$ empirical orthogonal function (EOF1) at the free end ($t=0$), computed over $110^{\circ}$--$50^{\circ}$~W, $15^{\circ}$--$50^{\circ}$~N and normalized by each model's largest absolute loading. The dashed box marks the track domain ($90^{\circ}$--$70^{\circ}$~W, $20^{\circ}$--$36^{\circ}$~N). Each model's PC1 sign is chosen so that higher PC1 means a more eastern free-end storm position and EOF1 is flipped to match, so red marks where heights are higher in members with high PC1. (Bottom) Storm central mean sea level pressure (MSLP, hPa) over local time (EDT). Rewind-AR retains 867 out of 1,000 after stability screening, and all 867 pass track quality control (QC). cBottle-video retains 947 after track QC (7 lack a qualifying Ian track and 46 have no storm detected at the free end). Gold lines for top and bottom rows mark the model members closest to the ERA5 reference for storm position (using frame-by-frame root-mean-square great-circle distance) and central pressure (using RMSE), respectively. ERA5 references are native-grid for Rewind-AR and HEALPix-round-tripped for cBottle-video.}
\label{fig:ian-tracks-intensity}
\end{figure}

Figure~\ref{fig:ian-tracks-intensity} compares 66-hour ensemble trajectories final-conditioned on ERA5 at the time of Hurricane Ian's South Carolina landfall (September 30, 2022, 18:00 UTC). In cBottle-video, the leading eight free-end $z_{500}$ principal components are only weakly associated with free-end storm position and intensity, explaining 9\%, 8\%, and 14\% of the in-sample variance (7\%, 6\%, and 12\% cross-validated) in free-end latitude, longitude, and central MSLP, respectively (Supplementary Figure~\ref{fig:si-eof-regression-ian}e--g), as reported by \citet{Lin2026-sz}. In contrast, the same regressions applied to Rewind-AR's 867 stability-screened members explain 71\%, 87\%, and 31\% in-sample and 70\%, 87\%, and 29\% cross-validated, as seen in Supplementary Figure~\ref{fig:si-eof-regression-ian}b--d. For Rewind-AR, PC1 alone orders antecedent tracks from west to east: low-PC1 members start over the eastern Gulf of Mexico and high-PC1 members over the Atlantic east of Florida (Figure~\ref{fig:ian-tracks-intensity}, top left). The corresponding EOF1 (Figure~\ref{fig:ian-tracks-intensity}, middle left) has a positive center over the Florida peninsula that extends northeastward along the U.S. East Coast, a compact negative center just east of Florida, and a broad negative center over the north-central United States. The compact pair of centers around Florida is what a storm displaced eastward produces in $z_{500}$ (lower heights where the storm moves to, higher heights where it would otherwise have been), so EOF1 likely records the storm's own position as well as any steering flow. By contrast, cBottle-video's EOF1 has a single broad center over the northeastern United States and southeastern Canada, far from the storm (Figure~\ref{fig:ian-tracks-intensity}, middle right), consistent with its weak association with storm position. The two models also are biased in opposite directions relative to ERA5 when it comes to hurricane intensity. Hurricanes generated by Rewind-AR generally start off more intense and weaken while the opposite is true for hurricanes generated by cBottle-video (as seen in Figure~\ref{fig:ian-tracks-intensity}, bottom; Supplementary Figure~\ref{fig:si-eof-regression-ian}d,g). Finally, ERA5's free-end storm lies at the southern edge of Rewind-AR's antecedent positions, whereas some cBottle-video storms begin even farther south (Figure~\ref{fig:ian-tracks-intensity}, top; Supplementary Figure~\ref{fig:si-eof-regression-ian}b,e).

\subsection{2021 Pacific Northwest heatwave}

Supplementary Figure~\ref{fig:pnw-free-end-states} compares 66-hour final-conditioned ensemble trajectories ending on June 28, 2021, at 12:00 UTC (05:00 PDT). Every cBottle-video member is warmer than its ERA5 reference at the free-end frame ($t=0$, June 25, 11:00 PDT), by $+0.44$ to $+5.8$\,K over land (Supplementary Figure~\ref{fig:pnw-free-end-states}j--p). The members then hold roughly steady while ERA5 strengthens past them into the June 27 peak, so cBottle-video misses the observed day-over-day warming (Supplementary Figure~\ref{fig:pnw-free-end-states}r).

In contrast, Rewind-AR's free-end states are mostly cooler than ERA5: only 2 of its 778 members are warmer (warmest $+0.2$\,K, coolest $-3.77$\,K; Supplementary Figure~\ref{fig:pnw-free-end-states}b--h), and the ensemble mean is 1.9\,K cooler. Toward the pinned final state, the ensemble captures the day-over-day warming into the June 27 peak, but its daytime heating is too weak: at 11:00 and 17:00 PDT on June 26 and 27, every member is cooler than ERA5 and the ensemble mean is 1.0--1.6\,K cooler, whereas at night it stays within 0.4\,K of ERA5 (Supplementary Figure~\ref{fig:pnw-free-end-states}q).

Antecedent surface temperatures in Rewind-AR are more strongly (though still weakly) associated with the large-scale height field---an expected finding given the central role of the blocking high in driving the heatwave \citep{Schumacher2022-gb}. An EOF analysis of the free-end $z_{500}$ anomaly field ($170^\circ$--$100^\circ$~W, $25^\circ$--$70^\circ$~N) reveals that the leading eight principal components explain 17\% of across-member land-mean $t_{2\text{m}}$ variance in-sample and 15\% under 10-fold cross-validation (within-model in-sample OLS F-test, $p < 0.001$; Supplementary Figure~\ref{fig:pnw-eof-regression}b). By contrast, the leading eight modes computed separately for cBottle-video explain 9\% in-sample and 8\% under cross-validation (Supplementary Figure~\ref{fig:pnw-eof-regression}c).

\section{Discussion and conclusion}

Backward autoregression learns the conditional antecedent distribution $p(\mathbf{x}_{t-\Delta t}\mid\mathbf{x}_t)$ rather than reversing physics, as coarse-graining removes the dissipation whose reversal would cause small errors to grow explosively \citep{Hassanzadeh2026-dd}. Within the 66-hour window, the leading $z_{500}$ modes explain far more of the across-member variance in Rewind-AR's ensembles than in cBottle-video's, Rewind-AR better follows the heatwave's observed day-over-day warming, and it keeps the 500~hPa ageostrophic wind fraction close to ERA5's (Supplementary Figure~\ref{fig:si-ageostrophic}). Several caveats apply. First, numerical stability screening already removes 13.3\% of Ian members and 22.2\% of Pacific Northwest members by 66 hours (and none of the Sandy members, a case inside both models' training periods), and all Rewind-AR statistics use only the surviving members. Second, while Rewind-AR shows a closer relationship between geopotential height EOFs and antecedent storm position, this alone does not imply higher physical plausibility. For example, for Superstorm Sandy, some hurricanes originate in implausible locations deeper inland (Supplementary Figure~\ref{fig:si-sandy-tracks-intensity}). Third, the two models also differ in resolution, training data, architecture, and size, so this does not represent a fully standardized comparison of autoregression and video diffusion for final-conditioned ensembles. Ultimately, more work is needed to ensure that final-conditioned ensembles created by AI weather and climate emulators represent plausible histories and are dynamically admissible. This work is ongoing.

\section*{Code Availability}
The code used to train Rewind-AR is available at
\url{https://github.com/jerrylin96/run-it-back}. The trained Rewind-AR
checkpoint, its normalization statistics and invariants, and the code that
loads it (\texttt{run-it-back} v1.0.0) are archived on Zenodo
\citep{Lin2026-ab}. Scripts used to compare
Rewind-AR with cBottle-video are available at
\url{https://github.com/jerrylin96/rewinding-the-extremes/tree/main/scripts/backward_emulator}.

\section*{Author Contributions}
Jerry Lin: Software, Conceptualization, Writing -- original draft, Writing --
review and editing. Elizabeth A. Barnes: Supervision, Funding acquisition,
Conceptualization, Writing -- review and editing.

\begin{ack}
J.L. was supported by the U.S. Department of Energy Office of Science through
the Program for Climate Model Diagnosis and Intercomparison (PCMDI).
We thank Zeyuan Hu (NVIDIA Corporation) for presenting the poster on Jerry
Lin's behalf at the Tackling Climate Change with Machine Learning workshop at
NeurIPS 2026 in Sydney, Australia.
\end{ack}

\section*{Use of Generative AI and AI-Assisted Technologies}
Claude Fable 5.1, Claude Fable 5, Claude Opus 4.8, and Claude Opus 5.5, accessed through Claude Code on the web, were used to edit code and text, assist with literature searches, and draft the initial
Supporting Information. GPT-5.6 Sol in OpenAI Codex was used for adversarial
code review. GPT-5.6 Sol in OpenAI Prism and Gemini 3.8 Flash in Antigravity
CLI were used to revise text for flow and clarity. Arena.ai was used to refine
overly complex sentences. GPT-6 Luna in OpenAI Prism was used to prepare these
camera-ready statements and edit the manuscript's LaTeX front matter. The
authors reviewed and verified the AI-assisted material and take responsibility
for the final manuscript.

\clearpage

{\small
\bibliographystyle{unsrtnat}
\bibliography{references}
}

\clearpage
\appendix
\section{Supplementary information}

\setcounter{figure}{0}
\renewcommand{\thefigure}{S\arabic{figure}}
\renewcommand{\theHfigure}{S\arabic{figure}}

\begin{figure}[ht]
  \centering
  \includegraphics[width=\linewidth]{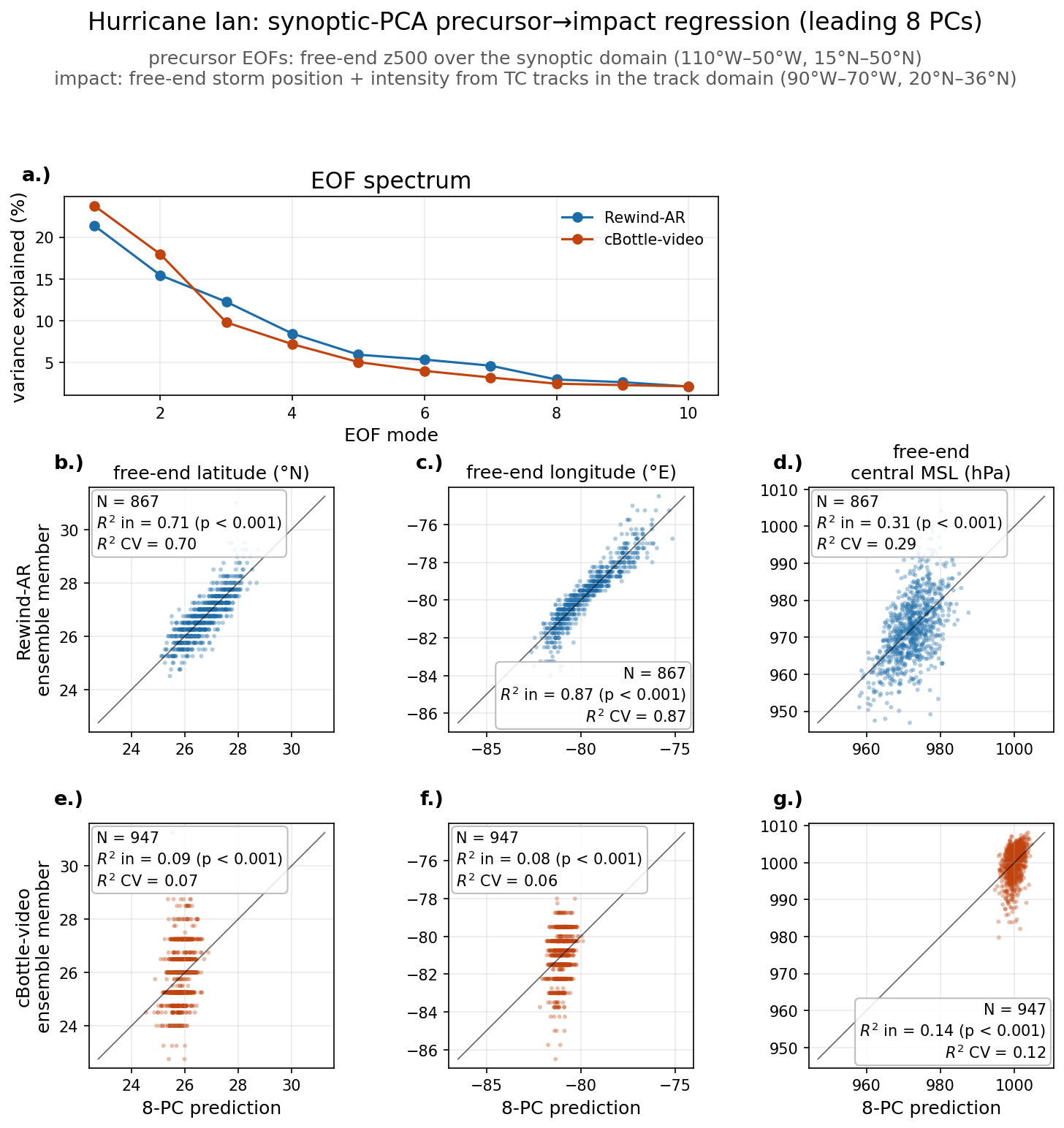}
  \caption{Hurricane Ian free-end synoptic EOF and regression diagnostics. (a) Variance explained by the leading EOFs of $z_{500}$ over $110^\circ$--$50^\circ$~W and $15^\circ$--$50^\circ$~N. (b--d) Rewind-AR member values ($N = 867$) versus predictions from regressions on that ensemble's leading eight principal components for free-end storm latitude, longitude, and central mean sea level pressure, respectively. (e--g) As in (b--d), for cBottle-video ($N = 947$), using its own leading eight principal components. Insets report in-sample/CV coefficients of determination; gray lines denote one-to-one agreement.}
  \label{fig:si-eof-regression-ian}
\end{figure}

\begin{figure}[t]
  \centering
  \includegraphics[width=\linewidth]{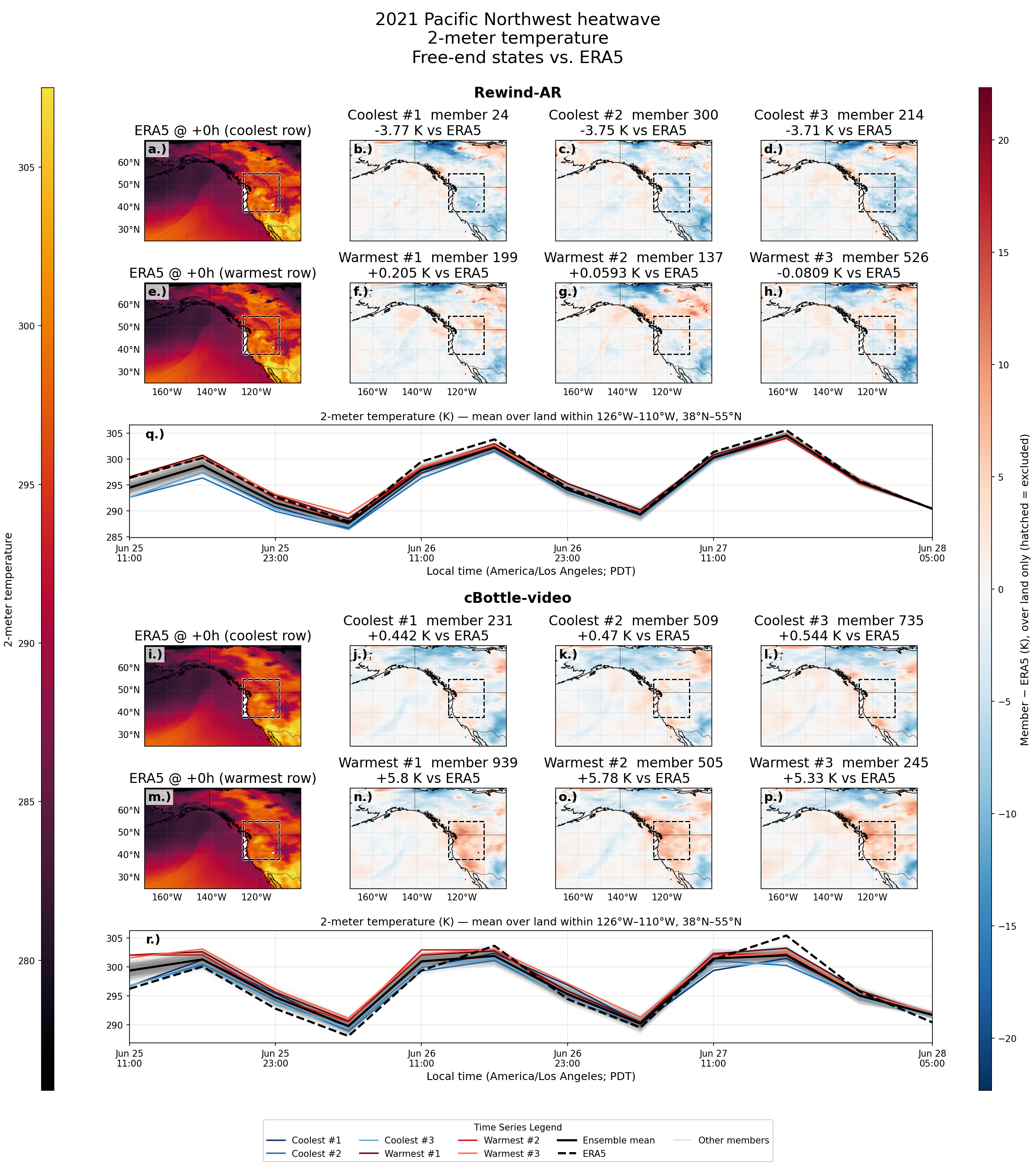}
  \caption{2021 Pacific Northwest heatwave 66-hour final-conditioned ensemble comparison between Rewind-AR and cBottle-video. (a--h) Free-end ($t=0$, June 25, 11:00 PDT) 2-meter temperature ($t_{2\text{m}}$) states and anomalies relative to ERA5 for Rewind-AR, showing the three coolest (b--d) and three warmest (f--h) ensemble members averaged over land within $126^\circ$--$110^\circ$~W and $38^\circ$--$55^\circ$~N (dashed box). (i--p) As in (a--h), but for cBottle-video. (q,r) Evolution of land-mean $t_{2\text{m}}$ (K) over local time (PDT) across ensemble members for Rewind-AR and cBottle-video, respectively, with dashed black lines denoting ERA5 verification. Rewind-AR retains 778 of its 1,000 members after stability screening. ERA5 references are native-grid for Rewind-AR and HEALPix-round-tripped for cBottle-video.}
  \label{fig:pnw-free-end-states}
\end{figure}

\begin{figure}[t]
  \centering
  \includegraphics[width=\linewidth]{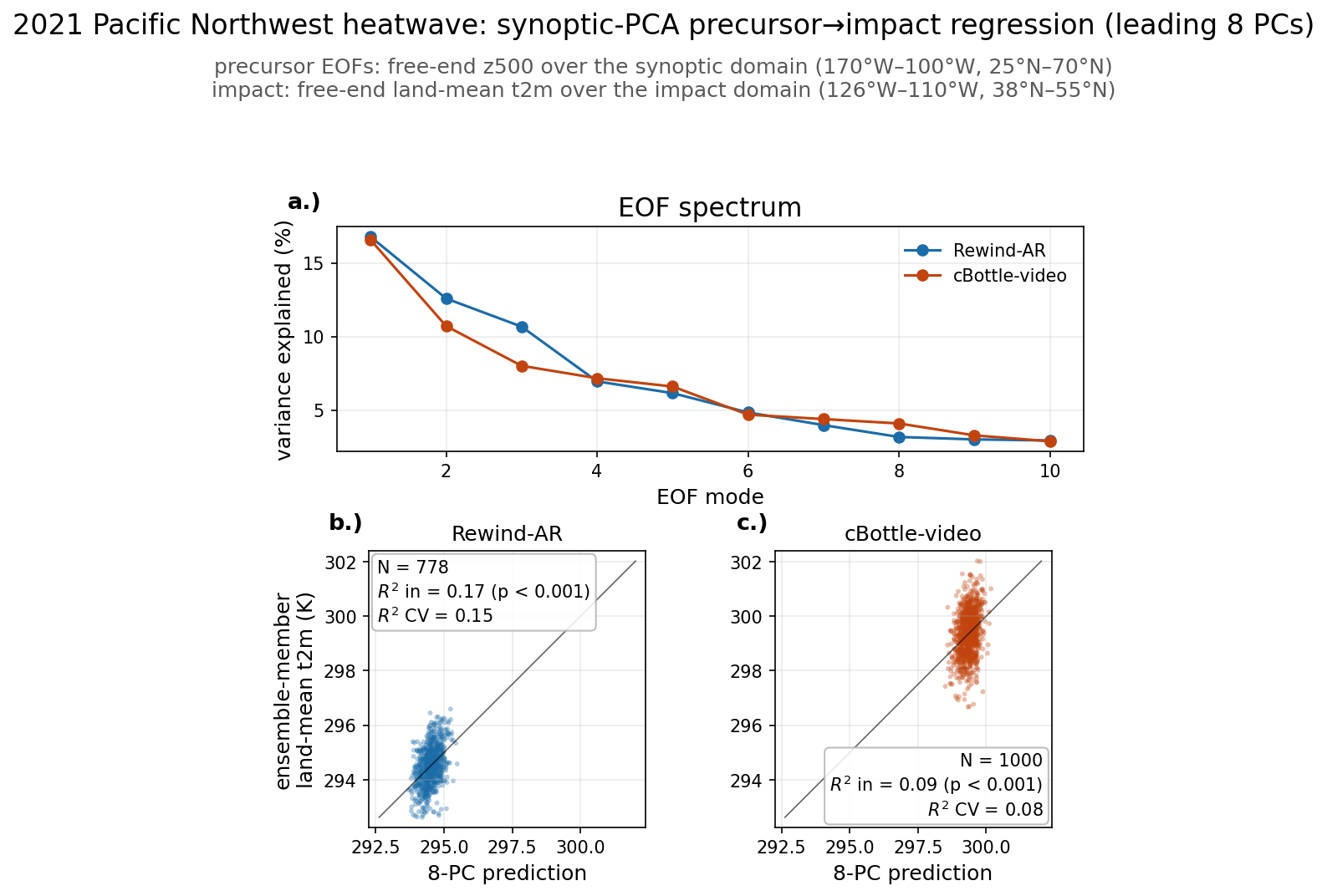}
  \caption{Pacific Northwest heatwave free-end synoptic EOF and regression diagnostics. (a) Variance explained by the leading EOFs of $z_{500}$ over $170^\circ$--$100^\circ$~W and $25^\circ$--$70^\circ$~N. (b,c) Ensemble-member land-mean $t_{2\text{m}}$ over $126^\circ$--$110^\circ$~W and $38^\circ$--$55^\circ$~N versus predictions from regressions on each ensemble's own leading eight principal components for Rewind-AR ($N = 778$) and cBottle-video ($N = 1{,}000$), respectively. Insets report in-sample/CV coefficients of determination; gray lines denote one-to-one agreement.}
  \label{fig:pnw-eof-regression}
\end{figure}

\begin{figure}[ht]
  \centering
  \includegraphics[width=\linewidth]{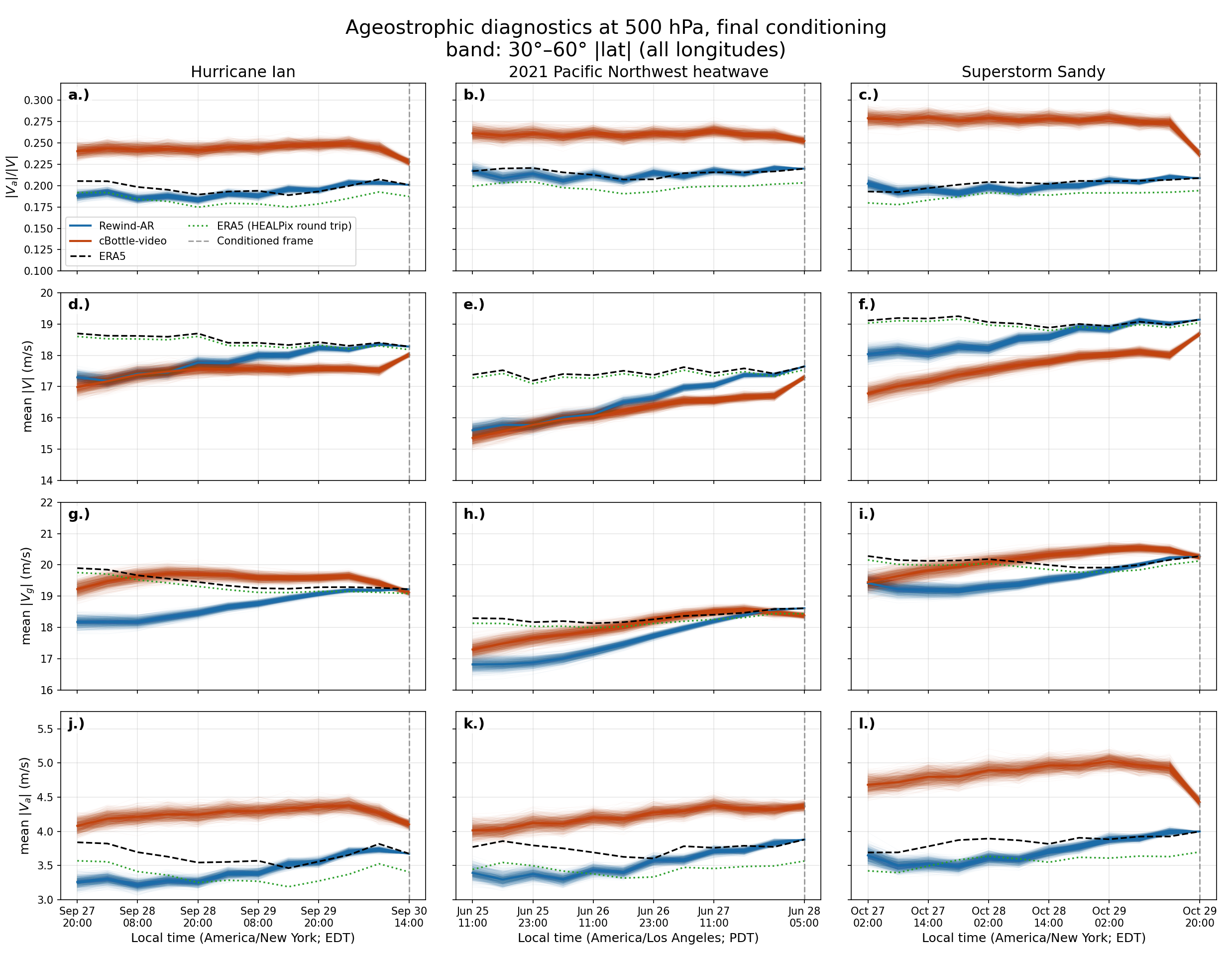}
  \caption{500~hPa ageostrophic diagnostics for the final-conditioned ensembles of Hurricane Ian (left), the 2021 Pacific Northwest heatwave (center), and Superstorm Sandy (right), averaged over $30^\circ$--$60^\circ$ latitude in both hemispheres. Rows show (a--c) the ageostrophic fraction $\langle|\mathbf{V}_a|\rangle/\langle|\mathbf{V}|\rangle$ and the mean speeds (d--f) $\langle|\mathbf{V}|\rangle$, (g--i) $\langle|\mathbf{V}_g|\rangle$, and (j--l) $\langle|\mathbf{V}_a|\rangle$. Thin lines are members and bold lines ensemble means: Rewind-AR in blue ($N = 867$, 778, and 1,000 stability-screened members) and cBottle-video in orange ($N = 1{,}000$). Rewind-AR's reference is native ERA5 (dashed black); cBottle-video's is ERA5 passed through the same HEALPix round trip as its conditioning data (dotted green). Gray dashed lines mark the pinned final frame.}
  \label{fig:si-ageostrophic}
\end{figure}

\begin{figure}[ht]
  \centering
  \includegraphics[width=\linewidth]{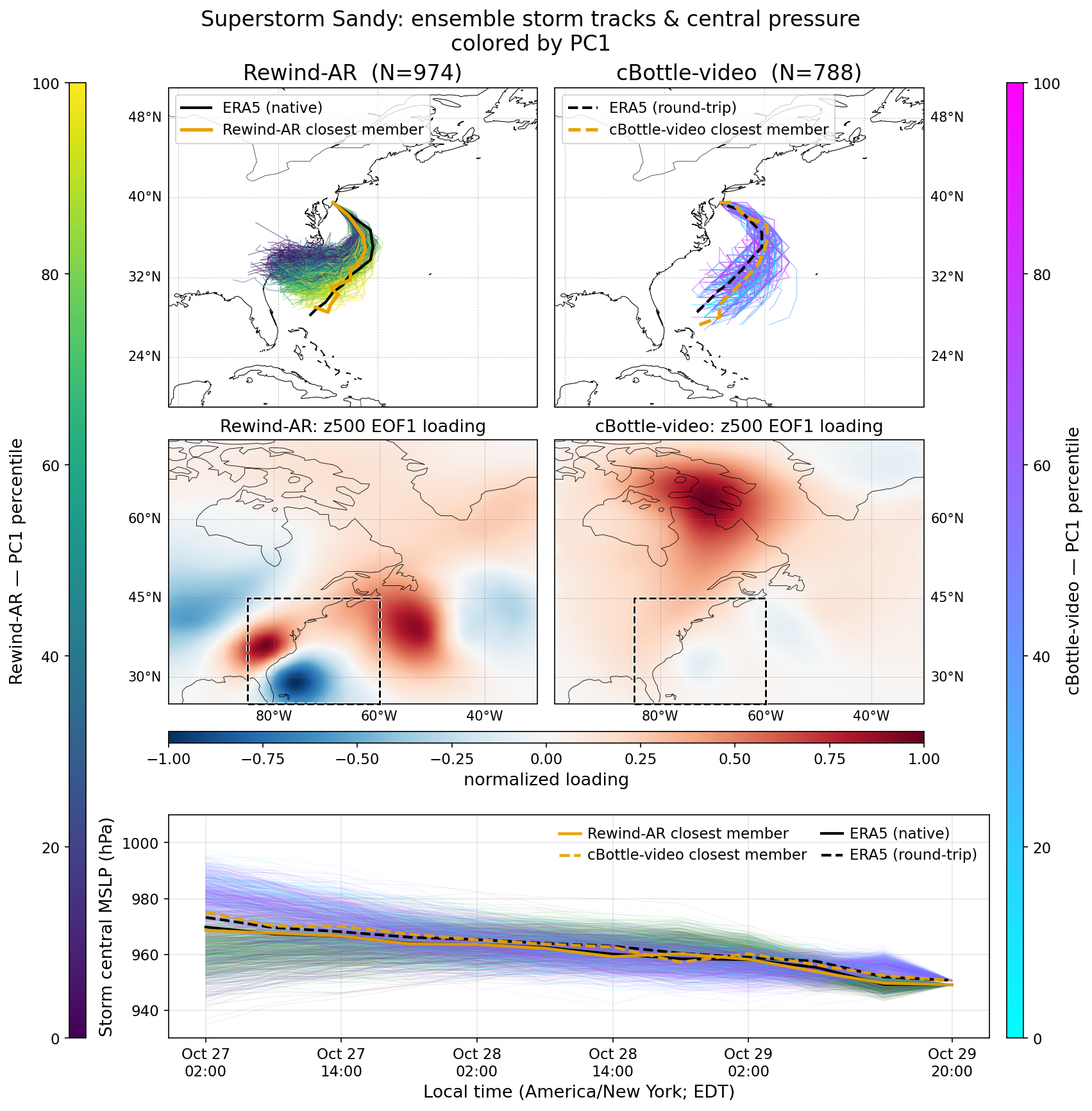}
  \caption{As in Figure~\ref{fig:ian-tracks-intensity}, for Superstorm Sandy: tracks within the track domain ($85^{\circ}$--$60^{\circ}$~W, $25^{\circ}$--$45^{\circ}$~N), EOF1 of $z_{500}$ over $100^{\circ}$--$30^{\circ}$~W, $25^{\circ}$--$75^{\circ}$~N, and central MSLP in EDT. 974 of Rewind-AR's 1,000 members and 788 of cBottle-video's pass track QC.}
  \label{fig:si-sandy-tracks-intensity}
\end{figure}

\begin{figure}[ht]
  \centering
  \includegraphics[width=\linewidth]{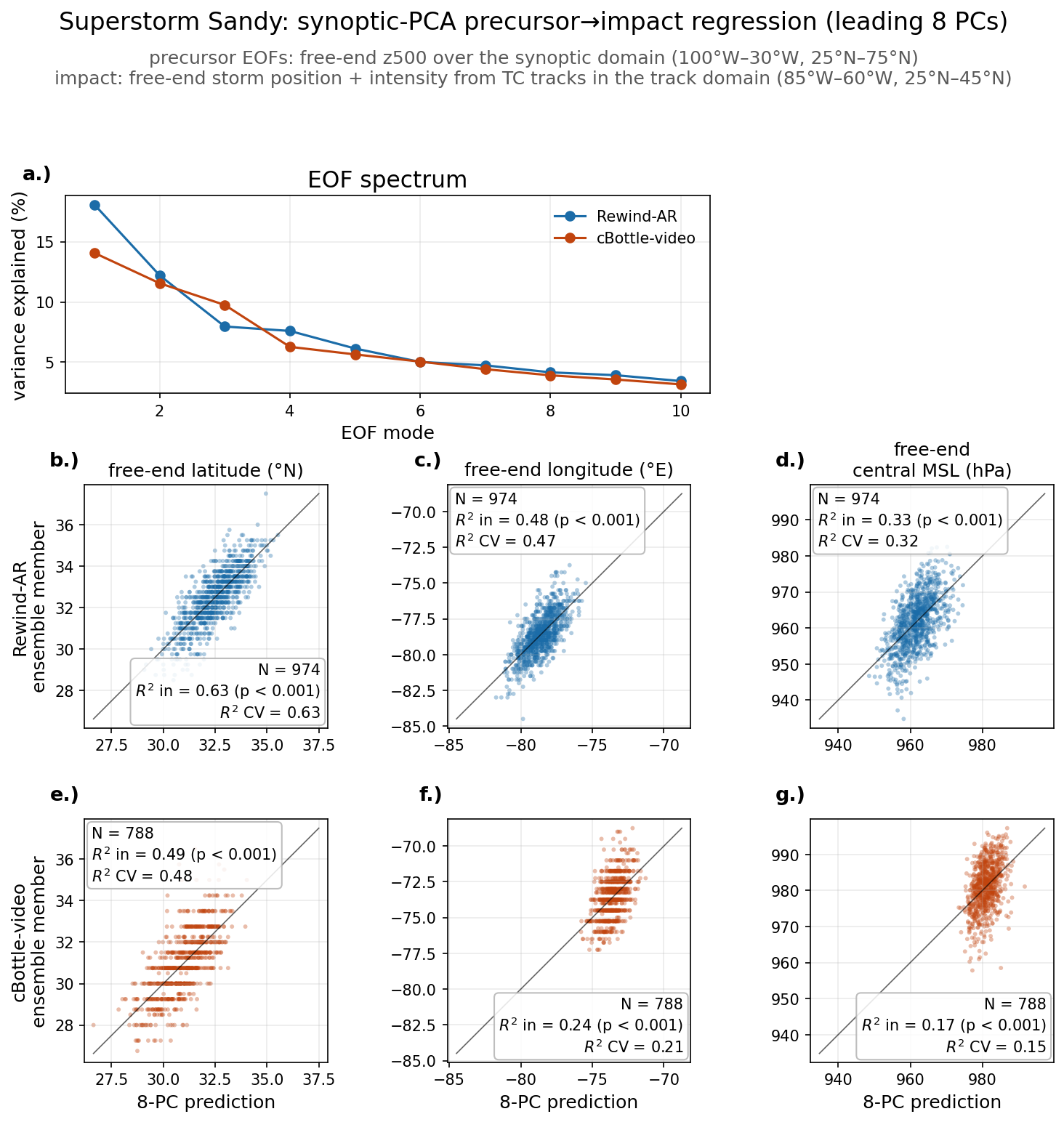}
  \caption{As in Supplementary Figure~\ref{fig:si-eof-regression-ian}, for Superstorm Sandy (Rewind-AR $N = 974$; cBottle-video $N = 788$), with EOFs of $z_{500}$ over $100^\circ$--$30^\circ$~W and $25^\circ$--$75^\circ$~N.}
  \label{fig:si-eof-regression-sandy}
\end{figure}

\subsection{Case-study ensemble screening and regression diagnostics}
\label{sec:si-screening}

Each case and model started with 1,000 members. Rewind-AR retains 867 Ian members, 778 Pacific Northwest members, and 1,000 Sandy members after numerical stability screening, which excludes 13.3\%, 22.2\%, and 0\% of the respective ensembles by 66 hours. The screen rejects a member once its global $t_{2\text{m}}$ field contains a nonfinite value or a value outside $[150,350]$\,K, and the member stays rejected for the rest of the backward rollout. Because the screen keeps only members that remain numerically stable, the survivors may not be a random sample of the ensemble. Track quality control follows \citet{Lin2026-sz}. Each member's storm is the tracked path that passes within 500~km of ERA5's storm fix at the pinned frame ($\pm 2$ frames allowed), and the member is kept only if that path has at least three fixes inside the case's track domain as well as a fix at the free-end frame. All 867 stability-screened Rewind-AR Ian members and 974 of the 1,000 Rewind-AR Sandy members pass track quality control. cBottle-video is not stability-screened, so it keeps all 1,000 Pacific Northwest members, and it retains 947 Ian and 788 Sandy members after track quality control. For Ian, 7 lack a qualifying track, and in 46 the storm is first detected 6--12 hours after the free end, so they have no free-end storm position. For Sandy, 44 lack a qualifying Sandy track, and 168 have no free-end storm position.

For each case and model, the EOF basis is computed separately from that ensemble's free-end $z_{500}$ anomalies about its ensemble mean. For Rewind-AR, it is computed from its stability-screened members on the native $0.25^\circ$ grid with anomalies scaled by $\sqrt{\cos\phi}$ so that the covariance is area-weighted by $\cos\phi$, and for cBottle-video it is computed from all 1,000 members on equal-area HEALPix pixels \citep{Lin2026-sz}. The hurricane regressions use only the members that pass track quality control. The leading eight principal components are held fixed for 10-fold cross-validation: OLS is refit in each fold, but PCA is not. Reported $p$-values are within-model in-sample OLS F-tests and do not test differences between models.

\subsection{500~hPa ageostrophic fraction}

Following \citet{Lin2026-sz}, the geostrophic wind is computed from 500~hPa geopotential by centered finite differences on the $0.25^\circ$ latitude--longitude grid, and the ageostrophic wind is the residual $\mathbf{V}_a = \mathbf{V} - \mathbf{V}_g$. Speeds are averaged with $\cos\phi$ weights over $30^\circ$--$60^\circ$ latitude in both hemispheres, and the ageostrophic fraction is the ratio of band means, $\langle|\mathbf{V}_a|\rangle/\langle|\mathbf{V}|\rangle$. cBottle-video fields are regridded from HEALPix to the $0.25^\circ$ grid before the calculation while Rewind-AR fields are used on their native grid. Averaged over the 11 unconditioned frames, Rewind-AR's ensemble-mean fraction differs from native ERA5 by $-0.005$ (Ian), $-0.002$ (Pacific Northwest), and $-0.002$ (Sandy), and by at most 0.017 at any frame, whereas cBottle-video's exceeds its round-tripped ERA5 reference by $+0.06$, $+0.06$, and $+0.09$ (Supplementary Figure~\ref{fig:si-ageostrophic}a--c). The departure starts at the conditioning frame: cBottle-video's fraction is already above its reference at the conditioned frame, which it regenerates rather than copies, and rises further at the first unconditioned frame, most sharply for Sandy and Ian, whereas Rewind-AR's matches ERA5 at the conditioned frame by construction and stays close to it. Rewind-AR therefore keeps the 500~hPa flow as close to geostrophic balance as ERA5. Its winds nonetheless weaken toward the free end: there, its mean total and geostrophic speeds fall 1.1--1.8 and 0.9--1.7\,m\,s$^{-1}$ below ERA5's (Supplementary Figure~\ref{fig:si-ageostrophic}d--i), so its height gradients weaken along with its winds. cBottle-video's total speed falls 1.6--2.2\,m\,s$^{-1}$ below its reference, but its geostrophic speed only 0.5--0.8\,m\,s$^{-1}$, which is what raises its ageostrophic fraction.

\subsection{Superstorm Sandy}

Superstorm Sandy (2012) falls inside both models' training periods (Rewind-AR: 2009--2018; cBottle-video: 1980--2017), unlike the 2021 Pacific Northwest heatwave and Hurricane Ian, so this case does not test generalization to an unseen event. The window ends on October 30, 2012, at 00:00 UTC, near Sandy's New Jersey landfall. All 1,000 Rewind-AR members pass stability screening, and 974 pass track QC. cBottle-video retains 788 members after track QC \citep{Lin2026-sz}. The leading eight free-end $z_{500}$ principal components explain more of the across-member variance in free-end storm latitude, longitude, and central MSLP for Rewind-AR (63\%, 47\%, and 32\% cross-validated) than for cBottle-video (48\%, 21\%, and 15\%; Supplementary Figure~\ref{fig:si-eof-regression-sandy}), though by a smaller margin than for Ian. Some of Rewind-AR's low-PC1 storms start unrealistically over land in the southeastern United States. As for Ian, Rewind-AR's EOF1 has a compact pair of centers around the storm, consistent with the storm's own displacement, whereas cBottle-video's EOF1 is a broad center over northeastern Canada (Supplementary Figure~\ref{fig:si-sandy-tracks-intensity}). ERA5's free-end storm again lies at the southern edge of Rewind-AR's members, whereas some cBottle-video storms start farther south than its reference, and most Rewind-AR storms start deeper than ERA5's, whereas most cBottle-video storms start shallower than its round-tripped reference.

\clearpage

\subsection{Rewind-AR}
\label{sec:methods_fcn3_backward}

\paragraph{Formulation and backbone architecture.}
To reconstruct antecedent atmospheric states, we formulate a single-step backward autoregressive model $f_\theta: \mathbf{x}_t \mapsto \mathbf{x}_{t - \Delta t}$ with time step $\Delta t
  = 6\,\text{h}$. Because forward atmospheric dissipation renders the time-reversed mapping inherently one-to-many, deterministic regression collapses unresolvable fine scales
  toward a blurry conditional mean. We therefore employ the FourCastNet~3 (FCN3) architecture configured as a probabilistic ensemble model \citep{Bonev2025-dw}. The model backbone consists of a 10-layer spherical neural operator (273,507,540 real scalar parameters, counting each complex weight as two real scalars; approximately $273.5\times 10^6$ parameters, latent width $d=384$, latent spatial downsampling factor of $2$) in which each block combines a pointwise MLP with a spherical convolution---local discrete-continuous (DISCO) convolutions in most blocks, and a global spectral convolution (in the manner of SFNO) in the first block of every five (2 of 10 blocks).

\paragraph{State representation and auxiliary conditioning.}
The prognostic atmospheric state $\mathbf{x}_t \in \mathbb{R}^{72 \times 721 \times 1440}$ is defined on a $0.25^\circ$ equiangular grid from ERA5 reanalysis, comprising 72 vertical and surface channels: 5 upper-air fields ($u, v, z, T, q$) across 13 pressure levels (50--1000\,hPa) and 7 surface fields ($u_{10\text{m}}, v_{10\text{m}}, u_{100\text{m}}, v_{100\text{m}}, t_{2\text{m}}, \text{MSL}, \text{TCWV}$). Static and dynamic auxiliary inputs are mapped through a dedicated auxiliary encoder ($d_{\text{aux}}=36$): surface geopotential ($Z$), land-sea mask ($\text{LSM}$), input-aligned top-of-atmosphere solar zenith angle $\cos\theta_z(t)$, and 8 multi-scale, spatially correlated stochastic noise channels, each generated by a diffusion process on the sphere with spatial scale set by $k_T \in [3.08\times 10^{-5}, 5.05\times 10^{-1}]$. Wind vector components $(u, v)$ are normalized via wind-angle-aware standardization (shared variance per level based on total horizontal wind magnitude), moisture fields ($q, \text{TCWV}$) are min-max scaled with a non-negative output clamp, and all remaining fields use standard $z$-score normalization.

\paragraph{Objective and training setup.}
The model is trained on 10 years of ERA5 reanalysis (2009--2018; validated on 2019) using an ensemble continuous ranked probability score (standard CDF-based CRPS) with a spectral penalty:
\begin{equation}
\mathcal{L} = \sum_{c=1}^{72} w_c \left[ \mathcal{L}_{\text{CRPS}}\!\left(\hat{\mathbf{x}}_{t-\Delta t}^{(c)}, \mathbf{x}_{t-\Delta t}^{(c)}\right) + 0.1\,
\mathcal{L}_{\text{Spectral-CRPS}}\!\left(\hat{\mathbf{x}}_{t-\Delta t}^{(c)}, \mathbf{x}_{t-\Delta t}^{(c)}\right) \right],
\end{equation}
where the raw channel weights $a_c$ are $0.1$ for surface fields, $a_{t_{2\text{m}}} = 1.0$, and $a_c = 0.001 \times p\,[\text{hPa}]$ for atmospheric pressure levels, and $w_c = a_c / \sum_j a_j$ are their normalized values. Both loss terms additionally include the channelwise factor $s_c / \max(\sigma_{\Delta x,c}, 10^{-4})$, where $s_c$ is the state-normalization scale and $\sigma_{\Delta x,c}$ is the six-hour tendency standard deviation; this factor may be absorbed into the loss terms, so final effective multipliers need not sum to one. Spectral-CRPS is CRPS on magnitudes of spherical-harmonic coefficients. Optimization uses Adam ($\text{lr} = 5\times 10^{-4}$, $\beta_1 = 0.9, \beta_2 = 0.95$, gradient clipping norm $16$,
global batch size $4$, training ensemble size $E=8$) with a \texttt{ReduceLROnPlateau} scheduler. Multistep backward rollouts are computed autoregressively: step $k$ predicts $\hat{\mathbf{x}}_{t-(k+1)\Delta t}$ from $\hat{\mathbf{x}}_{t-k\Delta t}$ conditioned on $\cos\theta_z(t-k\Delta t)$ with the noise channels advanced autoregressively from step to step.

State-normalization and six-hour tendency statistics were computed over 1979--2018 and reused unchanged. We evaluated the epoch-14 checkpoint (iteration 51,128), selected by minimum validation loss for the combined CRPS and spectral-CRPS objective. Rollout checks during training on 12 initial conditions from 2019 showed that validation loss stopped tracking rollout quality after epoch 7. CRPS at $-60$ hours was lowest at epoch 7, and rollouts to $-120$ hours, already numerically unstable by then, were worse at epochs 10 and 13. Unfortunately, the epoch-7 weights were not retained.

\subsection{cBottle-video}
\label{sec:methods_cbottle}

\paragraph{Architecture and spherical grid discretization.}
cBottle-video \citep{Brenowitz2025-mz} employs a continuous-time score-based diffusion model using an adapted Song-UNet architecture \citep{Song2020-hs, Karras2022-xq}. The network incorporates learned position embeddings, diurnal and seasonal time embeddings, AMIP mid-month SST boundary conditions linearly interpolated to each
frame's valid time, and temporal self-attention layers across frames \citep{Ho2022-lf, Blattmann2023-zo}. To circumvent spherical distortion and pole singularities in conventional latitude-longitude grids, the model operates on an equal-area HEALPix (HPX64) spherical discretization with custom face padding \citep{Karlbauer2024-nv}. HPX64 partitions the globe into 12 square faces of $64 \times 64 = 4{,}096$ pixels (49,152 total pixels globally), yielding an effective spatial resolution of $\sim$100\,km.

\paragraph{Masked video diffusion and score-based sampling.}
Training employs a masked-conditional video diffusion (MCVD) framework \citep{voleti2022MCVD} over 66-hour sequences (12 frames at $\Delta t = 6\,\text{h}$). During training, stochastic masking patterns (random frame dropout, block-wise masking, endpoint interpolation, and unconditional masking) enable arbitrary temporal inpainting
\citep{Brenowitz2025-mz}. The training objective is:
\begin{equation}
\label{eq:cbottle_training_objective}
\begin{aligned}
\min_{\theta}\; &\mathbb{E}_{\mathbf{x},m,\sigma,\epsilon}\left[\lambda(\sigma)
\left\|\mathcal{D}_\theta(\mathbf{x} + \epsilon; \sigma, \mathbf{c}_m) - \mathbf{x}\right\|^2\right],\\
&\lambda(\sigma) = \frac{\sigma^2 + \sigma_{\text{data}}^2}{\sigma^2\sigma_{\text{data}}^2},
\qquad \epsilon \sim \mathcal{N}(0, \sigma^2 \mathbf{I}),
\end{aligned}
\end{equation}
where $\mathbf{c}_m$ denotes the temporal-mask and auxiliary conditioning associated with mask $m$, and $\sigma_{\text{data}} = 1$. The training noise level $\sigma$ is sampled log-uniformly over $[0.02, 1000]$. We evaluate a single public cBottle-video denoiser at all sampling noise levels. Samples are drawn across 18 discretization steps by integrating the reverse-time probability-flow ODE:
\begin{equation}
\label{eq:cbottle_sampling_step}
\frac{d\mathbf{x}}{d\sigma} = \frac{\mathbf{x} - \mathcal{D}_\theta(\mathbf{x}; \sigma, \mathbf{c}_m)}{\sigma}
\end{equation}
using the deterministic second-order Heun sampler of \citet{Karras2022-xq} with $S_{\text{churn}} = 0$.

\paragraph{Conditioning and inference setup.}
We evaluate the public pre-trained checkpoint \citep{Brenowitz2025-mz} via \texttt{Earth2Studio} \citep{Earth2Studio_Contributors_NVIDIA_Earth2Studio_2024}, implementing
flexible masked conditioning by subclassing its inference wrapper. Conditioning frames are extracted from $0.25^\circ$ ERA5 reanalysis \citep{Hersbach2020-oh} and regridded to HPX64. ERA5 at $t=66\,\text{h}$ supplies final-frame conditioning, and the sampler jointly generates all 12 frames without overwriting its generated endpoint with ERA5, for each of the 1,000 ensemble members.

\end{document}